\documentclass [12pt,epsfig]{article}
\usepackage{amsmath,amssymb,epsfig}
\usepackage{graphicx}

\begin{document}
\newcommand{\el}{\left}
\newcommand{\er}{\right}
\newcommand{\dis}{\displaystyle}
\newcommand{\p}{\prime}
\newcommand{\ka}{\kappa}
\newcommand{\bc}{\beta C}
\newcommand{\ro}{\rho^\circ}
\newcommand{\ti}{\tilde}
%
%

\begin{center}
{\bf\Large Testing $^{6,8}$He density distributions by
calculations of \\
total reaction cross-sections of $^{6,8}$He+$^{28}$Si }\\ [.3cm]

{ V.K. LUKYANOV, E.V. ZEMLYANAYA } \\
 {\footnotesize\it Joint
Institute for Nuclear Research, Dubna 141980, Russia }\\[.1cm]

{ S.E. MASSEN, Ch.C. MOUSTAKIDIS} \\
 {\footnotesize\it Aristotle
University of Thessaloniki,  Greece } \\[.1cm]

{ A.N. ANTONOV} \\
{\footnotesize\it Institute for Nuclear Research and Nuclear
Energy,  Sofia 1784, Bulgaria  } \\[.1cm]

{  G.Z. KRUMOVA}\\
 {\footnotesize\it University of Rousse, Rousse
7017, Bulgaria}
\end{center}
%
%
\begin{abstract} Calculations of the $^{6,8}$He$~+~^{28}$Si total
reaction cross sections at intermediate energies are performed on
the basis of the Glauber-Sitenko microscopic optical-limit model.
The target-nucleus density distribution is taken from the
electron-nucleus scattering data, and the $^{6,8}$He densities are
used as they are derived in different models. The results of the
calculations are compared with the existing experimental data. The
effects of the density tails of the projectile nuclei as well as
the role of shell admixtures and short-range correlations are
analyzed.
\end{abstract}

\begin{sloppypar}

\section{Introduction}
\setcounter {equation}0
\vspace*{-0.5pt} \noindent The study of the matter distributions
in the Borromean nucleus $^6He$ and the neutron rich isotope
$^8He$ is an actual problem discussed in many papers (see, e.g.,
in \cite{conf} and refs. therein). Investigations give a reason to
expect long tails (halo) of the neutron densities. The first
result on the enhancement of the strong interaction radii $R_I$
for $^{6,8}He$ was deduced in \cite{Tan1} from measurements of
their interaction cross sections with different target nuclei at
fixed energy 800 MeV/nucleon. The same conclusion was made in
\cite{Tan2} about the $^{6,8}$He $rms$-radii. Later in
\cite{Tan3}, the model densities in the form of a sum of two
Gaussian functions \cite{Zhuk} were fitted to the data and a
conclusion for a significant neutron halo in the outer region of
these nuclei was made. As to the differential elastic scattering
cross sections, they were studied in collisions of $^{6,8}$He with
protons, and the matter distributions of these nuclei were tested
using different phenomenological and theoretical models. Elastic
scattering angular distributions of $^6He+p$, $^6He+{^4He}$
($E_{lab}=151\, MeV$) and $^6He+p$, $^8He+p$ ($E < 100 A\, MeV$)
were calculated in \cite{A6} and \cite{A7}, respectively, by means
of the real part of the optical potentials obtained
microscopically by using the realistic M3Y-Paris effective
interaction \cite{{A8},{A9},{A10}} and taking into account the
neutron and proton densities of helium isotopes from \cite{Tan3}
and also derived from the cluster-orbital shell-model
approximation (COSMA) \cite{Zhuk},\cite{A11},\cite{A12},\cite{Z1}.
It was shown in \cite{A7} that elastic scattering is a good tool
to distinguish between different density distributions. The
results of \cite{A7} were compared also with those of the
alpha-core approach with the fully nonlocal effective interactions
\cite{A13} and the no-core model based on the large space
shell-model calculations (LSSM) \cite{A14},\cite{A15},\cite{A4}.
Elastic scattering was recently studied also in \cite{Alkh}.

At the same time, small experimental information on the energy
dependence of the total reaction cross sections of $^{6,8}He$ with
nuclei is available. Generally, from such data for the well-known
densities of the target-nuclei, one can get additional constraints
on the parameters of the tested models of the halo nuclei. Below,
we present our calculations of the total reaction cross-sections
and compare them with  the respective data on interactions of
$^{6,8}He$ with $^{28}Si$ at energies E/nucleon $\sim$ 10$\div$ 50
MeV \cite{{swed},{Peny}}.


\section {Some formulae}

\noindent Following  \cite{LZS} we start with the expression for
the total reaction cross-sections in the framework of the
optical-limit model of the high-energy multiple-scattering theory
\cite{{G},{S}}:
\\
\begin{equation}\label{2.1}
\sigma_R=2\pi\int\limits_0^\infty db~b~\el(1- {\Huge e}^{-\dis
\chi(b)}\er).
\end{equation}
Here, the eikonal phase for the scattering of the projectile
nucleus ($p$) by the target nucleus ($t$) is given in \cite{Czyz}
by the expression
\\
\begin{equation}\label{2.2}
\chi(b)={\bar\sigma}_{NN}\,\int d^2s_p\,
d^2s_t~\ro_p(s_p)~\ro_t(s_t)~f(\xi), \qquad\quad {\bf\xi}=\bf
b-{\bf s}_p+{\bf s}_t.
\end{equation}
\\
For the isospin averaged total nucleon-nucleon cross-section
${\bar\sigma}_{NN}$ we use the parametrization from \cite{CG}. The
vectors $~{\bf s},~{\bf\xi}$ are in the impact parameter $b$ plane
perpendicular to the $z$- axis along the relative momentum ${\bf
k}_i$ of the two scattered nuclei. The thickness densities are
given as
\\
\begin{equation}\label{2.3}
\ro(s)=\int\limits_{-\infty}^\infty dz~\ro(\sqrt{s^2+z^2}),
\end{equation}
where the bare density distributions $\ro(r)$ of point nucleons in
respective nuclei depend on their coordinates in the
center-of-mass frames. The function $f(\xi)$ is defined by the
form of the nucleon-nucleon interaction amplitude.

To present the phase integral in one-dimensional form, one uses
the two-dimensional Fourier-Bessel transformation of all functions
in the integrand
\begin{equation}\label{2.4}
u(s)={1\over (2\pi)^2}\int \, e^{-\dis{i{\bf q}{\bf s}}} \,{\ti
u}(q)d^2q = {1\over 2\pi}\int\limits_0^\infty \, J_0(qs) \, {\ti
u}(q)qdq,
\end{equation}
where
\begin{equation}\label{2.5}
{\ti u}(q)= \int \, e^{\dis{i{\bf q}{\bf s}}} \,u(s)d^2s = 2\pi
\int\limits_0^\infty \, J_0(qs)\, u(s)sds.
\end{equation}
This results in:
\begin{equation}\label{2.6}
\chi(b)={\bar\sigma}_{NN}\,{1\over 2\pi}\int\limits_0^\infty
qdq~J_0(qb)~ {\ti \ro}_p(q)~{\ti \ro}_t(q)~{\ti f}(q).
\end{equation}
The original high-energy scattering theory uses only the
transversal component $q_{\perp}=q\sin{\vartheta/2}$ of the total
momentum transfer $q=2k\sin{\vartheta/2}$ and neglects its
longitudinal component $q_{||}=q\sin{\vartheta/2}$. Here
$q^2=q_{\perp}^2+q_{||}^2$, and $k$ is the momentum of the
relative motion. So, the theory is applied to the scattering
angles $\vartheta < \sqrt{2/kR}$, where $R$ is approximately the
sum of the radii of both the nuclei$\footnote {~For estimations of
the validity of this approximation one can see \cite{LZ}.}$. So,
in eq.(\ref{2.6}) the use of $q_{\perp}=q$ allows one to take the
3-dimensional Fourier transforms of the functions $\ro(r)$ and
$f(\xi)$ instead of the 2-dimensional transforms of their profile
functions $\ro(s)$ and $f(s)$. We remind that the form factors of
the point $\ti\ro(q)$ and nuclear $\ti\rho(q)$ densities obey the
relation
\begin{equation}\label{2.7}
\ti\rho(q)=\ti\rho_N(q)\,\ti\ro (q),
\end{equation}
where the nucleon form factor $\ti\rho_N(q)$ is suggested to be
the proton one and has a dipole form which is transformed to the
Gaussian function at small $q$:
\begin{equation}\label{2.8}
{\ti\rho}_N(q)\,=\,\Bigl(1+\frac{q^2
r^2_{0\,rms}}{12}\Bigr)^{-2}\simeq \exp(-q^2 r^2_{0\,rms}/6).
\end{equation}
At the same time the form ${\ti f}(q)$ of the NN-scattering
amplitude is known to be
\begin{equation}\label{2.9}
{\ti f}(q)\,=\,\exp(-q^2 r^2_{N\,rms}/6).
\end{equation}
The squared values of the $rms$ radii of the nucleon and the
NN-interaction, known from the experimental data as
$r^2_{0\,rms}=0.658\,fm^2$ and $r^2_{N\,rms}=0.66\pm 0.03\,fm^2$,
almost coincide. It was concluded in \cite{LZS} that a small
difference between these $rms$ radii does not influence visibly
the calculated total reaction cross sections, and, thus, one can
use approximately  $\ti\rho_N(q)={\ti f}(q)$. Then taking into
account Eqs. (\ref{2.7})-(\ref{2.9}) we get the phase (\ref{2.6})
in a simple form
\begin{equation}\label{2.10}
\chi(b)={\bar\sigma}_{NN}\,{1\over 2\pi}\int\limits_0^\infty
qdq~J_0(qb)~ {\ro}_p(q)~{\rho}_t(q).
\end{equation}
The form factors ${\rho}_t(q)$ can be calculated by using the data
on nuclear density distributions. Here, we use the realistic
density in the form of the symmetrized Fermi density distribution
\begin{equation}\label{2.11}
\rho(r)=\rho_0\, {\sinh R/a\over \cosh R/a \,+\, \cosh r/a}\simeq
\rho_0\,{1\over 1+\exp[(r-R)/a]},
\end{equation}
and the respective form factor obtained in \cite{LPP} (see also
\cite{GKLS}) in the form
\begin{equation}\label{2.12}
\ti\rho(q)=-\rho_0\,{4\pi a\over q}\,{d\over dq} \el[{\sin
qR\over\sinh\pi aq}\er], \qquad \rho_0={3A\over 4\pi
R^3}\,\el[1+\el({\pi a\over R}\er)^2\er]^{-1}.
\end{equation}
Here $R$ and $a$ are the radius and diffuseness parameters,
respectively. Now, taking different models for the projectile
nuclei we can test them by comparison of the calculated cross
sections with the existing experimental data. Note that if for the
densities of the projectile and the target nucleus one takes the
Gaussian functions and uses (\ref{2.9}) for $f(q)$, then the
integration in (\ref{2.6}) can be performed explicitly \cite{LZS},
and one obtains the result of \cite{Ka} frequently used by
experimentalists analyzing experimental data
\begin{equation}\label{2.13}
\chi(b)={\bar\sigma}_{NN}{1\over
\pi}\frac{A_pA_t}{a^{\circ~2}_{G,p}+
a^{\circ~2}_{G,t}+a_N^2}\,\exp \Bigl(-{b^2\over
a^{\circ~2}_p+a^{\circ~2}_t+a_N^2}\Bigr),
\end{equation}
where $a_N^2=(2/3)r^2_{N\,rms}$. In the case of the zero-range
approximation $a_N=0$. For the repulsive Coulomb field the
trajectory of the incident nucleus deflects from the scattering
center. This effect can be taken into account  replacing he impact
parameter $b$ in the phase $\chi(b)$ t by the distance of the
closest approach $b_c$ in the Coulomb field
\begin{equation}\label{2.14}
 b \,\rightarrow \, b_c={\bar a} + \sqrt{{\bar a}^2+b^2},
\end{equation}
\noindent where ${\bar a} = Z_pZ_t e^2/2E_{c.m.}$ is the
half-distance of the closest approach at $b=0$.

In our work, we use models for the projectile nuclei which are
known in a table form and, thus, we apply expression (\ref{2.10})
in the calculations where the density $\rho_t(q)$ of the target
nucleus ${^{28}Si}$ is taken in a realistic form (\ref{2.11}) with
$R=3.085\,fm$ and $a=0.563\,fm$ from \cite{BKLP}. The Coulomb
distortion effect is also taken into account by using
(\ref{2.14}).

\section{Density distributions and calculated cross section}

\noindent We apply the model for the s-p and s-d shell nuclei,
developed in Ref. \cite{MM} (M-model), to calculate the
${}^{6,8}$He densities. The densities of these projectile nuclei
are used to study the effect of the Jastrow-type short range
NN-correlations (SRC), on the total reaction cross-section
${}^{6,8}$He $+ {}^{28}$Si measured at different energies
\cite{swed,Peny}.

In the M-model, a general expression for the one-body density
matrix of $N=Z$, $s$-$p$ and $s$-$d$ shell nuclei \cite{MM} has
been obtained using the factor cluster expansion method (see
Ref.\cite{Clark} and Refs. therein). That expression depends on
the harmonic oscillator (HO) parameter $B$
($B=(\hbar/(m\omega))^{1/2}$), the occupation numbers $A_{nl}$ of
the various states and the correlation parameter $\beta$ that
comes from the Jastrow type correlation function
\begin{equation}\label{3.1}
 f(r_{ij})=1-\exp[-\beta (r_i - r_j)^2], \quad r_{ij}=|{\bf r_i}-{\bf
r_j}|
  \end{equation}
which introduces short range correlations. The effect of
correlations introduced by the function $f(r)$ becomes large when
the correlation parameter $\beta$ becomes small and vice versa.

The expression of $\rho({\bf r})$ (in the two body approximation
for the cluster expansion)  has the form
\begin{equation}\label{3.2}
\rho(r)=N \left[ \rho_{SD}^o(r_B) + O_2(r_B,\beta) \right],\quad
r_B=r/B
 \end{equation}
where  $\rho_{SD}^o(r_B)$, the uncorrelated density distribution
corresponding  to the Slater determinant HO total wave function,
comes from the one-body term of the cluster expansion of the
one-body density matrix and the term $O_2(r_B,\beta)$ comes from
the two-body term. Their expressions as well as the expression of
the normalization factor $N$ are given in \cite{MM}. Similar
expressions for the form factor and the momentum distribution have
also been given in Ref. \cite{MM}.

The parameters $B$, $\beta$ and $A_{nl}$ can be determined, for
each nucleus, by fit to the experimental form factors and momentum
distributions. As such data are not available for ${}^{6,8}$He we
will use the data of the total reaction cross-section and the
$rms$-radius. That is, we involve the parameters $B$, $\beta$ and
the four occupation number $A_{1s}$, $A_{1p}$, $A_{1d}$ and
$A_{2s}$ in the fitting procedure.

We used as experimental data the well known results of Tanihata et
al \cite{Tan3} (T-model) for the density distributions obtained by
fit the total reaction cross-section of ${}^{6,8}$He $+{}^{12}$C
at 800 MeV in calculation with the model discussed in Sec. 1,
using for the target nucleus the Gaussian form $\rho_G(r)$ fitted
to the realistic density in the outer region and using for the
projectile nucleus the harmonic-oscillator form
\begin{equation}\label{3.3}
\rho_X^o (r)=\frac{2}{ ( {\tilde a}\sqrt{\pi})^3 }\, e^{(r/
{\tilde a})^2} + \frac{X-2}{3}\frac{2}{ ({\tilde b}\sqrt{\pi})^3
}\frac{r}{\tilde b}\,
  e^{(r/ {\tilde b})^2},
\end{equation}
where $X=Z,N$, and $\tilde a$, $\tilde b$ are the fitting
parameters within the T-model.

Expression (\ref{3.3}) and the corresponding expression of the
$rms$-radius have been used as the "experimental" data of the
projectile nuclei ${}^{6,8}$He. The values of the parameters $B$,
$\beta$ and $A_{nl}$ which have been found by a least squares fit
of the theoretical neutron and proton density distributions
(expression (\ref{3.2})) to the "experimental" ones with the
constrain that the two densities have the same $rms$-radius are
given in Table I. In the case of proton densities there is one
free parameter, the HO parameter. For that reason the proton
density is taken to be the same for both models.

\begin{figure}[htbp]
\begin{center}
\epsfig{file=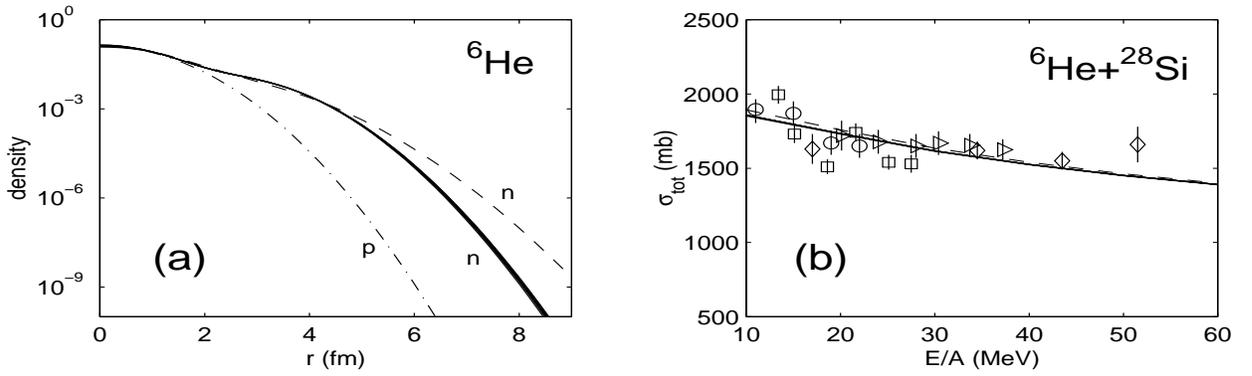,height=5cm,width=1.\linewidth}
\end{center}
\caption{ (a) Point density distributions of He-6 calculated using
the M-model \cite{A4}. Dashed line: neutron density; dash-dotted
line: proton density; bold curve is a bundle of eight curves (see
the explanations in the text); (b) Total reaction cross section
for He-6 + Si-28. The curves are as in a).}
\end{figure}
%
\begin{figure}[htbp]
\begin{center}
\epsfig{file=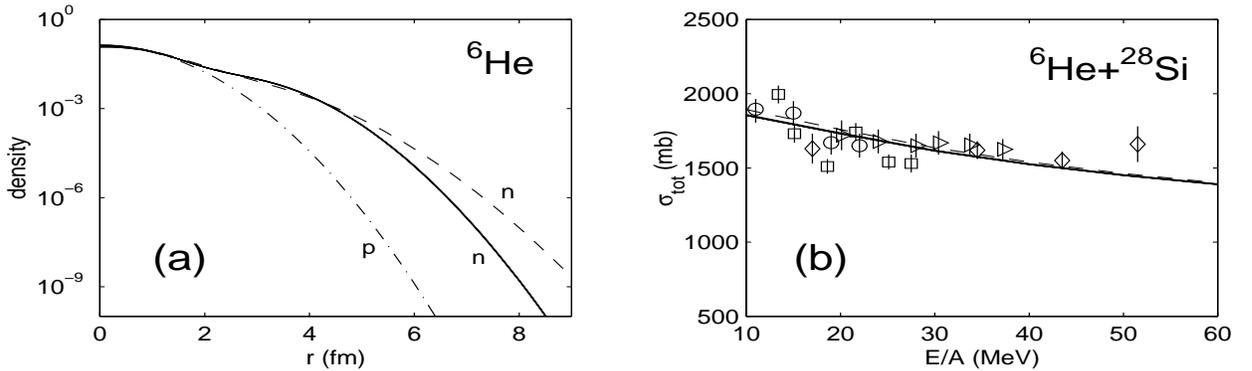,height=5cm,width=1.\linewidth}
\end{center}
\caption{ The same as in Fig. 1 from calculations within the
M-model \cite{A4} where only the parameter beta was used in the
fitting procedure. }
\end{figure}

In order to study the effect of SRC on the density and the total
cross section we examined two cases. In the first case we have
taken the neutron occupation numbers to be the ones given in Table
I, we have given to the correlation parameter $\beta$ various
values from $\beta=4\, $fm$^{-2}$ to $\beta=50\, $fm$^{-2}$ and
for each value of $\beta$ we determined the parameter $B$ so that
the neutron $rms$-radius to be the same with the one obtained
within the T-model. In figure 1 we present the point density
distribution for ${}^{6}$He calculated in this case. The bold
curve is the bundle of eight figures. It is seen that the neutron
densities of the M-model have lower tails at $r>5\,$fm compared to
the one of the T-model. At the same time, we see that the
cross-sections are in good coincidence. This means that the main
contribution comes from the region $r=0\div 5\,$fm, i.e., up to
$r$ equal to about two $rms$ radius of the ${}^{6}$He nucleus (the
$rms$-radius of the mater distribution is 2.331 fm for both
model). From a physical point of view, the increase of the SRC at
small $\beta$ enhances the nuclear radius, but in the considered
case one can think that this effect is compensated by reducing the
HO parameter $B$ which results in "squeezing" the wave functions
of the basis. For that reason we considered the second case where
the calculations were made for various values of the parameter
$\beta$ in the range $4$ to $50\,$fm$^{-2}$ while the HO parameter
$B$ had the value 1.8249 fm and the parameters $A_{nl}$ were taken
as before from Table I. The results are presented in Fig. 2. The
calculations in Ref.\cite{MM} showed that the changes of the
values of $\beta$ in that region give very significant changes in
the form factors and the momentum distributions of nuclei. In the
present paper, we study the total cross-sections and as a result,
we obtained the corresponding spread of the $rms$ radius of the
neutron density for ${}^{6}$He in the limit $2.6030\div
2.5757\,$fm. However, this does not noticeably influence the
cross-sections. Once again, the cross-sections of the M- and the
T-models are in coincidence, while the $rms$ radii are slightly
different.

The same conclusions can be made for ${}^{8}$He nucleus from Figs.
3 and 4, where the results of the calculations within the M-model
are presented. In Fig. 3, one can see the bunch of nine curves for
the neutron densities when the pairs of the parameters $\beta$ and
$B$ which give the $rms$-radius $R_{rms,n}=2.6587\,$fm are in the
regions $\beta=3.1588\div 50\,$fm$^{-2}$ and $B=1.7850\div
1.8261\,$fm. The occupation numbers are the ones given in Table I.
For all the neutron densities the $rms$-radii are in coincidence
with the T-model $R_{rms,n}=2.6857\,$fm. The matter density radius
is also the same in both models $R_{rms,A}=2.6857\,$fm. The values
of the parameters of the curves in Fig. 4 are: $\beta=3.1558\div
10.1558\,$fm$^{-2}$, $B=1.8075\,$fm and $A_{nl}$ are the same as
in Fig. 3. The obtained values of the $rms$ radii of the neutron
densities are in the region $2.7172\div2.6676\,$fm. We see that in
both figures the slopes of the densities from the M-model begin to
differ from that of the T-model at about $5\,$fm, where the
densities fall down by about two orders of magnitude comparing to
the values in the nuclear center. Their difference in the range of
twice the $rms$-radius does not affect the total cross-sections.

\begin{table}
\caption{The values of the parameters $B$, $\beta$ and $A_{nl}$ ,
for the neutron and proton densities of  ${}^{6,8}$He, obtained by
fit of the expression (\ref{3.2}) to expression (\ref{3.3}) with
the constrain that the two densities to give the same
$rms$-radius.} \vspace*{.1cm}
\begin{center}
\begin{tabular}{l c c c c c c c}
\hline
 Nucleus&  $B$[fm]   &  $\beta$[fm]${}^{-2}$&  $A_{1s}$&
  $A_{1p}$&  $A_{1d}$&  $A_{2s}$&  $R_{rms}$[fm]\\
\hline
 ${}^6$He(n) &1.8249& 6.0001& 1.9999& 1.0889& 0.&0.91124 &2.5865 \\
 ${}^8$He(n) &1.8075& 5.1558& 2.0000& 2.9056& 0.& 1.0944 &2.6587\\
${}^6$He(p)  &1.53  & $\infty$& 2.& 0.& 0.&0. &1.72 \\
 ${}^8$He(p) &1.53  & $\infty$& 2.& 0.& 0.&0. &1.76 \\
 \hline
 \end{tabular}
\end{center}
 \end{table}
%
%
\begin{figure}
\begin{center}
\epsfig{file=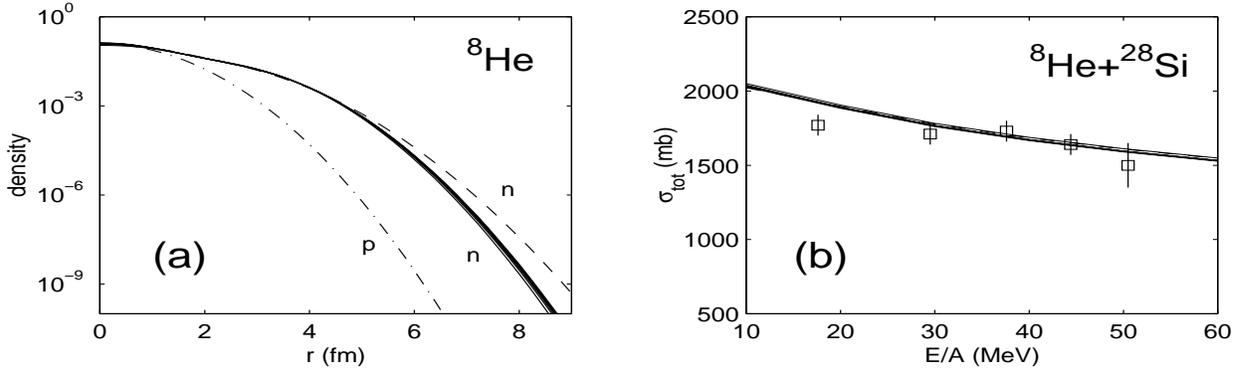,height=5cm,width=1.\linewidth}
\end{center}
\caption{ (a) Point density distributions of He-8 calculated using
the M-model [17]. Dashed line: neutron density; dash-dotted line:
proton density; bold curve is a bundle of neutron densities; (b)
Total reaction cross section for He-8+Si-28. }
\end{figure}

\section{Discussion and conclusions}

\noindent One should mention that in the M-model there were no
calculations of the binding energies and the structure of excited
states of these nuclei. However, large-scale shell-model
calculations (LSSM) of the wave functions and characteristics of
light exotic nuclei, such as ${^{6,8}He}$ and ${^{9,11}Li}$, were
reported by several groups, e.g. \cite{{A4},{A1},{A2},{A3},{A5}}.
They are usually performed by using interactions obtained directly
from NN g-matrices with various NN-interactions as their base. In
\cite{A5}, the wave functions for ${^{6,8}He}$ were calculated
within a complete 4$\hbar\omega$ model space by using the g-matrix
with interaction from \cite{A61}. All calculations were performed
by the shell-model code OXBASH \cite{A71}. For $^4He$ the
harmonic-oscillator single-particle wave functions were used,
while for ${^{6,8}He}$ they were the Woods-Saxon ones. It was
concluded in \cite{A5} that $^6He$ corresponds to halo nucleus,
while $^8He$ is a neutron skin nucleus. In \cite{{Zhuk},{Z1}} (and
refs. therein), the K-harmonic method was developed and
calculations were based on the assumption of the $1p_{3/2}$ state
for each of the valence neutrons related to the alpha-core center
(COSMA-model). The admixtures of the 3-body forces is included to
get the proper value of the binding energy.

\begin{figure}[htbp]
\begin{center}
\epsfig{file=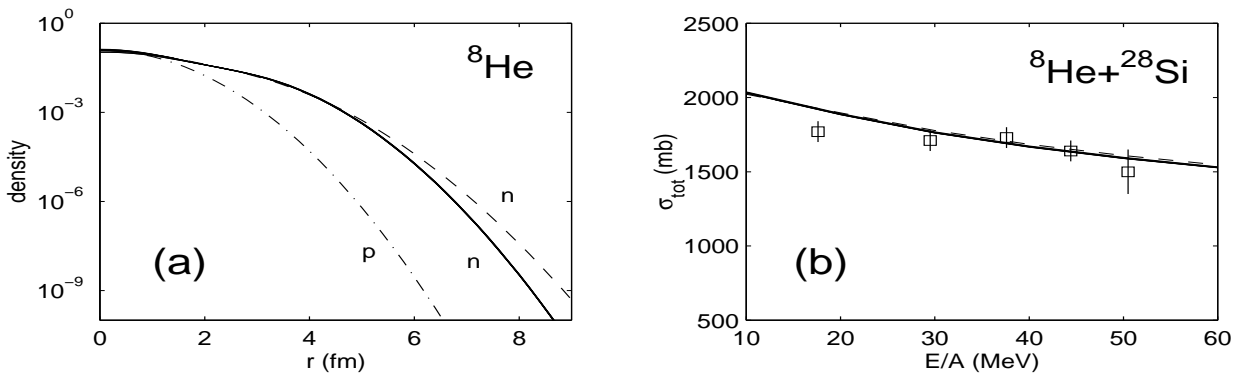,height=5cm,width=1.\linewidth}
\end{center}
\caption{ The same as in Fig.3 with differences in the fitting
procedure (see the explanations in the text). }
\end{figure}
%

In Fig.5, we show the matter density distributions for $^6He$
obtained from the calculations in the framework of the COSMA-model
\cite{Z1} (dash-dots) and the LSSM-model \cite{A5} (dots). The M-
and T-model densities are shown by the solid bunch and the dashed
curve, respectively. First of all, one can see that in the outer
region the COSMA- and LSSM-model densities have extended tails
which are followed by exponential asymptotics. Here, the dotted
curve exceeds the dash-dotted one which results in the enhancement
of the corresponding cross section. It is seen that both the cross
sections are about $200\div 500\,mb$ above those calculated by the
M- and T-models and the experimental data. The respective $rms$
matter radii are $R_{rms,\,N} =2.560\,fm$ (for the dash-dotted
curve) and  $R_{rms,\,N}=2.956\,fm$ (for the dotted curve). In
this connection we refer to \cite{AkTT} where the coupling of
elastic and elastic breakup channels were shown to play an
important role in processes with weakly bound nuclei, and this can
diminish the cross section obtained without accounting for the
additional channels. Recently, the theory of the coupled channels
was developed in \cite{{E1},{E2}} to study the breakup reactions
of halo nuclei. In general, the implied optical-limit model needs
higher order NN-scattering terms to be included as had been done
in the multiple-scattering theory \cite{{G},{S}}. For example, it
was shown in \cite{Alkh} that these terms provide a negative
interference and thus diminish the $p+{^6He}$ differential elastic
cross section calculated by taking into account only the
single-scattering term. The in-medium effect, which changes the
free NN-scattering cross section ${\bar\sigma}_{NN}$ in the phase
(\ref{2.2}) can be included as well. As shown qualitatively in
\cite{LZS}, this effect can decrease the total nucleus-nucleus
reaction cross sections by about 10\%. All the mentioned problems
regarding the mechanism of the nucleus-nucleus interaction are
still under investigations of the current theory. However, our
study of the total reaction cross-sections of the $^{6,8}He$
interactions with the ${^{28}Si}$ target leads to some general
conclusions which, we hope, will not be affected by a possible
further modification of the theory.

\begin{figure}[htbp]
\begin{center}
\epsfig{file=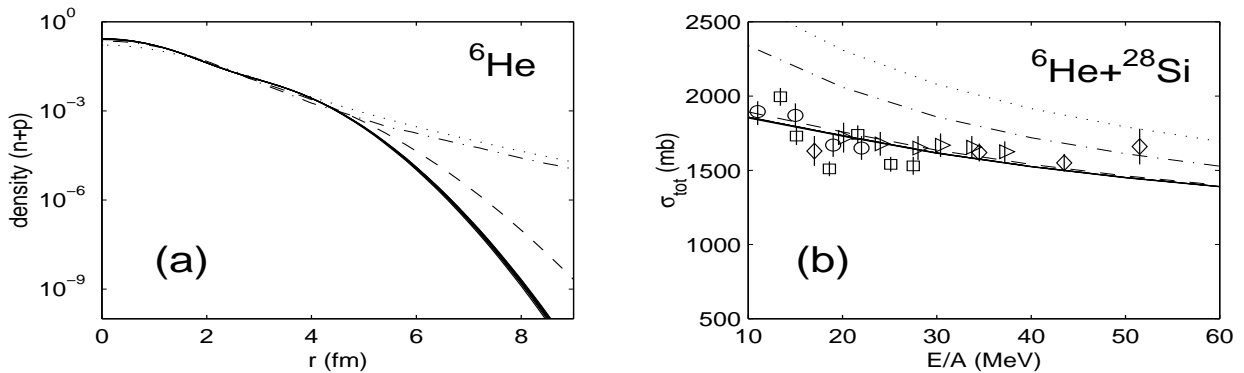,height=5cm,width=1.\linewidth}
\end{center}
\caption{  (a) Matter density distributions for He-6 calculated
with the COSMA model [21,22] (dash-dotted line) and the LSSM [27]
(dotted line). The M- and T-model densities are given by solid
bunch and dashed lines,respectively. (b) Total reaction cross
section for He-6+ Si-28. The curves are as in (a).}
\end{figure}

The following conclusions can be drawn from our results:
\begin{itemize}
\item  The $rms$ radii are more sensitive nuclear characteristics than the
total reaction cross section with respect to changes of the
density distributions. The radius depends strongly on variations
of the density in the nuclear interior while the behaviour of the
total reaction cross-section depends mainly on the peculiarities
of the nuclear surface and weaker on the central nuclear region
where strong absorption takes place.
\item  The previous item explains why variations of the densities due to
SRC do not affect  the total cross sections while they are
significant in calculations of the form factors and of the
single-particle momentum distributions in nuclei at large momentum
transfers.
\item  The fitting procedure in our work showed that there is not a
$1d$-shell contribution to the wave functions of $^6He$ and
$^8He$.
\item  In the case of the harmonic-oscillator basis, the Gaussian
type of the wave function asymptotics does not affect
significantly the cross sections and the latter are formed by
taking into account the interior region of the projectile nucleus
up to about only two $rms$ nuclear radii. Otherwise, if one gets a
basis for realistic wave functions, which have an exponential
asymptotics, the region of the main contribution becomes larger
and achieves the value of about four and larger times the $rms$
nuclear radius.
\end{itemize}

{\bf Acknowledgments}\\

 The authors would like to thank Dr.S.Ershov for
providing the $^6He$ matter density distribution used in
investigations in \cite{{E1},{E2}}. E.V.Z. was supported by the
RFBR under grant No.03-01-00657.

\end{sloppypar}
\end{document}